# Experiment for Testing Special Relativity Theory

L. I. Govor, G. A. Kotel'nikov[*], E. A. Meleshko, and G. V. Yakovlev

Russian Research Centre Kurchatov Institute, Kurchatov Sq., 1, Moscow, 123182 Russia
[*]E-mail: kga@kiae.ru

**Abstract**—An experiment aimed at testing special relativity via a comparison of the velocity of a non matter particle (annihilation photon) with the velocity of the matter particle (Compton electron) produced by the second annihilation photon from the decay $^{22}Na(\beta^+)^{22}Ne$ is proposed.

## 1. INTRODUCTION

In 2005, the physics community commemorated the 100th anniversary of the creation of the theory of special relativity, one of the outstanding achievements of human intellect. Throughout this period, special relativity, which is one of the cornerstones of the physics world outlook, has been subjected to testing of ever increasing precision. At the end of the $20^{th}$ and the beginning of the $21^{th}$ century, interest in such investigations grew, which was likely a consequence of general advances in theoretical and experimental physics. Special relativity is based on the requirement that equations that describe the motion of particles and fields be invariant under space-time transformations of the Lorentz group. Therefore, a violation of the Lorentz invariance of these equations would entail the violation of special relativity. Several versions of the theory that, in some way or another, go beyond the Lorentz group have been constructed. Below, we will dwell on those of the m that triggered intensive experimental investigations.

Ritz's emission theory (or ballistic theory) of 1908 [1] is the most well known am o n g these. According to Ritz, Euclidean geometry is realized in nature and the speed of light is $3 \times 10^{10}$ cm/s only with respect to the source of emission [1]. We denote this quantity by $c_0$. In accordance with the Galilean theorem of addition of velocities, the velocity of light emitted by a moving source, **c**, is then the vector sum of the velocity of light emitted by the source at rest, $\mathbf{c}_0$, and the velocity of source motion, **v**; that is, $\mathbf{c} = \mathbf{c}_0 + \mathbf{v}$. In contrast to this, the velocity of light in special relativity must always be isotropic and independent of the emitter velocity, $c = c_0$. Until recently, it was precisely this distinction that was the subject of study in many experiments {comparison of the velocities of light emitted by the west and the east equatorial edge of the solar disk [2], comparison of the velocities of photons for different angles of their divergence under conditions of in-flight electron–positron annihilation [3], comparison of the velocities of photons emitted by a moving excited $^{12}C^*$ nucleus from the reaction $^{12}C(\gamma, \gamma')^{12}C^*$ and a $^{16}O^*$ nucleus at rest from the reaction $^{16}O(\gamma, \gamma')^{16}O^*$ [4], comparison of the velocities of photons from the decay of fast $\pi^0$ mesons for different angles of photon divergence [5], and measurement of the absolute values of the velocities of photons fro m the decay of 6 - GeV $\pi^0$ mesons [6]}. As a result, it was shown that Ritz's ballistic theory does not agree with experimental results. It was found that, to a high precision ($c = c_0 + kv$, $k = (-4 \pm 13) \times 10^{-5}$ [6]), the velocity of light is isotropic and, within the form of the Galilean theorem of addition of velocities, is independent of the emitter velocity. Further advances required the formulation of new theoretical ideas and proposals concerning their experimental realization.

In 1977, Mansouri and Sexl [7] showed that the homogeneous transformation $t = a(u)T + \varepsilon(u)x$ and $x = b(u)(X - uT)$ of spacetime admits the existence of a preferred reference frame $\Sigma$, where $T$ and $X$ are, respectively, the time and the spatial coordinate along the abscissa in this reference frame. A special role of this reference frame is that, in it, the velocity of light—we will also denote it by $c_0$—is isotropic, while the synchronization of clocks situated at different places occurs according to Einstein's rule. An arbitrary reference frame $K$ in which $t$ is time and $x$ is the abscissa is assumed to move at a velocity $u$ along the $X$ axis with respect to . The quantities $a(u)$, $b(u)$, and $(u)$ are kinematical parameters of spacetime transformations. In an arbitrary reference frame, the velocity of light becomes anisotropic and depends both on the velocity and on the position of the reference frame $K$ with respect to the preferred reference frame : $c(u, ) = c_0[1+( - -1/2)(u^2/c_0^2)\sin^2 +( - +1)(u^2/c_0^2)]$ [8]. Here, is the angle between the direction of light propagation and the direction of the velocity **u** and the parameters , , and are related to the kinematical parameters $a(u)$, $b(u)$, and $(u)$. From the theoretical point of view, the dependence $c = c(u, )$ means a violation of the Lorentz invariance of laws of nature. But if Lorentz invariance survives, $( - -1/2)$ and $( - +1)$ must be zero, and this may be a subject of experimental tests. In the same year 1977, an anisotropy of 3 K relic radiation was discovered at a level of $(3.5 \pm 0.6) \times 10^{-3}$ K [9]. The result of this observation was interpreted as a possible "new ether drift"—a manifestation of the Doppler effect because of the absolute motion of the Solar System at a velocity of $u_S = 390 \pm 60$ km/s with respect to a preferred cosmological reference frame toward the Leo constellation [9]. It seemed that the preferred reference frame [7] found its place in physics as the reference frame in which relic radiation is isotropic. The presence of such a reference frame entails profound physics consequences, which were discussed by Glashow [10]. In the region of Planck energies $E_P = (hc_0^5/G)^{1/2} \sim 10^{19}$ GeV (where $h$ is the Planck constant and $G$ gravitational constant [11]), it is natural to expect special effects: the neutral pion may prove to be stable against the decay $^0$ 2 , while vacuum Cherenkov radiation, the proton beta decay $p^+$ $n^0 + e^+ +$ , and the photon decay $+ ^0$ are admissible [10]. The signatures of such processes may be observed in cosmic rays of energy in the range $10^{18}$–$10^{20}$ eV [10]. The formula $c = c(u, )$ proved to be the most convenient for tests. The three parameters , , and , which appear in this formula, can be determined from three independent experiments: from measurements of the Doppler effect, $( - -1/2)$ from the Michelson experiment, and $( - +1)$ from the Kennedy–Thorndike experiment [12]. In the present-day implementation, two mutually orthogonal cryogenic optical resonators manufactured from sapphire and cooled to a temperature of 4.2 K [8, 12, 13] play the role of mirrors in the Michelson interferometer. This makes it possible to improve the sensitivity of measurements by several orders of magnitude. In relevant experiments, the resonator frequencies were compared for a long time (up to one year) as the Earth rotated about the Sun. It was found that $= (-1/2 \pm 1.7) \times 10^{-7}$, $= (1/2 \pm 2.1) \times 10^{-5}$, and $= (0 \pm 2.1) \times 10^{-5}$ [8]; $( - -1) = (-3.1 \pm 6.9) \times 10^{-7}$ [12]; and $( - -1/2) = (-2.2 \pm 1.5) \times 10^{-9}$ [13]. The corresponding anisotropy of the velocity of light is $_\phi c/c_0 = (4.8 \pm 5.3) \times 10^{-12}$ in [8] and $_\phi c/c_0 = (2.6 \pm 1.7) \times 10^{-15}$ in [13]. A modern testing of the Doppler effect was performed in [14]. The Doppler shift of the frequency of the two-level transition $^3S_1(F = 5/2)$ $^3P_2(F = 7/2)$ in $^7Li^+$ ions at the ion velocity of $V = 0.064c_0$ (13.3 MeV) was measured. This experiment confirmed the special-relativity prediction to within $2.2 \times 10^{-7}$. The resulting constraint on the parameter is $( + 1/2) < 2.2 \times 10^{-7}$ [14].

A different approach to the violation of special relativity postulates was proposed in 1989 by Kostelelecký and Samuel [15]. They showed that string theory admits a spontaneous breakdown of Lorentz symmetry (and, accordingly, of special relativity) at an early stage of the evolution of the Universe at about $t_P = l_P/c_0 \sim 10^{-43}$ s, where $l_P = (hG/c_0^3)^{1/2} \sim 10^{-33}$ cm is the Planck length scale [11]. From the point of view of the present-day Lorentz-invariant state of matter, traces of this breakdown should manifest themselves in the existence of moderately small relic fields singling out preferred directions in 3-space, which violate its isotropy. In subsequent investigations, Colladay and Kostelelecký [16] included the idea of a spontaneous breakdown of Lorentz invariance in the Standard Model and constructed a version of an extended standard model (Standard Model Extension also known as SME). The SME is compatible with the Standard Model and contains all possible interactions that can arise upon spontaneous breakdown of Lorentz symmetry [17]. Its coefficients may be interpreted as constant background tensor fields corresponding to relic fields in string theory [17]. As a result, the physical properties of a particle, such as its momentum and energy, will change slightly, depending on the orientation of the laboratory frame with respect to the directions of the relic fields [16, 17].

Owing to its generality, the SME makes it possible to obtain deeper insight into simple models featuring a violation of Lorentz invariance [7] and to explain its existence [17]. Since these predictions are obviously fundamental, they are thoroughly tested in experiments. A number of studies on the subject have been reported. By means of interferometric measurements on the basis of cryogenic sapphire resonators [13, 18, 19] and by scanning the Moon's surface with lasers [20], it was shown that the SME coefficients do not exceed values of about $10^{-15}$–$10^{-11}$ in the photon sector and values of about $10^{-11}$–$10^{-6}$ in the gravity sector; that is, the velocity of light and surrounding space are isotropic to a high precision. This gives sufficient grounds to consider different models that go beyond Lorentz invariance, but which preserve the isotropy of the velocity of light and 3-space.

## 2. FIVE DIMENSIONS OF THE SPACE OF EVENTS: ENERGY AND MOMENTUM

The planned experiment is based on a model that admits the existence of five-dimensional physics space of events that is characterized by the metric [21–23]

$$ds^2 = c^2 dt^2 - dx^2 - dy^2 - dz^2, \qquad (1)$$

where $c$ is the velocity of light belonging to a continuum of values between $c_0$ and . The coordinates of an event point in this space (which is denoted here by $V^5$) are determined by the set of five numbers. These are the values of the spatial variables $x$, $y$, and $z$; the time $t$; and the velocity of light $c$. In terms of the variables $(x^0, \mathbf{x}, x^5)$, $x^0 = ct$ and $x^5 = c$, it contains two Minkowski subspaces characterized by the metric $g_{\mu\nu} = (+,-,-,-)$, $\mu$, $= 0, 1, 2, 3$: $M^4_1$ on the hyperplane of invariant velocity of light $x^5 = c_0 = 3 \times 10^{10}$ cm/s and $M^4_2$, where the time $dt' = dt$ is invariant and where the velocity of light $c$ is given by

$$c = c_0 \sqrt{1 + v^2/c_0^2}, \qquad (2)$$

with $0 \quad v < $ being the velocity of the source of light (emitter). Since the square of the velocity $v$ appears in (2), $c$ is independent of the direction of source motion and is therefore isotropic by construction. It follows that the experimental results reported in [2–6, 8, 12–14, 18–20] are compatible with the kinematics of motions in the $M^4_2$ space. In the case of free motions, the corresponding homogeneous integral transformations preserving the invariance of the interval in (1) can be represented in a form that admits a universal time

$$x' = \frac{x - Vt}{\sqrt{1 - V^2/c^2}}, \quad y' = y, \quad z' = z, \quad t' = t, \quad c' = \frac{c - Vx/ct}{\sqrt{1 - V^2/c^2}}, \qquad (3)$$

where $V$ is the velocity of the reference frame $K'$ with respect to the reference frame $K$ in $M^4_2$. In the $V^5$ space, the transformations in (3) are hyperbolic rotations $x' = x \cosh \quad - ct \sinh \quad$, $c' = -(x/t) \sinh \quad + c \cosh \quad$, and $\tanh \quad = \quad = V/c$ in the $(x, c)$ plane at a constant time $t' = t$ and form a one-dimensional Lorentz group $L_1$. This distinguishes them from the hyperbolic rotations $x' = x \cosh \quad - c_0 t \sinh \quad$, $t' = -(x/c_0) \sinh \quad + t \cosh \quad$, and $\tanh \quad = \quad = W/c_0$ ($W$ is the velocity of the reference frame $K'$ with respect to $K$ in $M^4_1$) in the $(x, t)$ plane within special relativity [24]. The Lorentz group $L_6$ at a constant time can be similarly defined in the general case. We specify its elements in terms of the six operators $N_{0i}$ and $N_{ik}$ belonging to the set $N_{0i} = ct \quad_i + (x_i/t) \quad_c$, $N_{ik} = x_i \quad_k - x_k \quad_i$, $P_0 = \quad/c$, $Q_0 = \quad/t$, $Q_i = \quad_i$, and $Z = c \quad_c - t \quad_t$, $i, k = 1, 2, 3$, $x^i = (x, y, z)$, $\quad_i = \quad/x^i$. In the set of functions $\quad = \quad(ct, \mathbf{x}) \subset f(t, \mathbf{x}, c)$, to which we restrict our consideration, they form the Lie algebra of the 12-dimensional group $(P_{10}, T_1) \times \quad_1$ [21–23]:

$$[Q_\mu, Q_\nu] = 0; \quad [Q_\mu, N_{\nu\rho}] = g_{\mu\nu} Q_\rho - g_{\mu\rho} Q_\nu; \quad [N_{\mu\nu}, N_{\rho\sigma}] = -g_{\mu\rho} N_{\nu\sigma} + g_{\mu\sigma} N_{\nu\rho} + g_{\nu\rho} N_{\mu\sigma} - g_{\nu\sigma} N_{\mu\rho}; \qquad (4)$$

$[P_0,Q_\nu]=0$; $[P_0,N_{\nu\rho}]=g_{0\nu}P_\rho - g_{0\rho}P_\nu$; $[Z,Q_\mu]=[Z,P_0]=[Z,N_{\mu\nu}]=0$.

Here, the set of operators $N_{0i}$, $N_{ik}$, $Q_0$, and $Q_i$ induces the inhomogeneous Lorentz group (Poincaré group) ₁₀; the operator $Q_0$ induces translations along the $c$ axis at constant $t$; $P_0$ induces translations along the $t$ axis at constant $c$ (translation group $T_1$); and the operator $Z$ induces the group ₁ of velocity of light–time scale transformations $c't' = ct$. The generators $N_{0i}$, $N_{ik}$, and $Z$ are the operators of symmetry of the surface $c^2t^2 - x^2 - y^2 - z^2 = 0$, which, at $c = c_0$, reduces to the light cone $c_0^2t^2 - x^2 - y^2 - z^2 = 0$. Under the corresponding space–time–velocity of light transformations, the equations of these surfaces go over to themselves.

The Lagrangian $L$ invariant under transformations induced by the algebra specified in (4) has the form $L = -m_0\sqrt{1-u^2} + (e/c_0)(A \cdot u - \varphi)$ [21–23], where $m$ is the mass of a particle, $\mathbf{u} = \mathbf{v}/c$ is the dimensionless velocity, $e$ is the electric charge, and $(\mathbf{A}, \varphi)$ are the components of the 4-potential. The generalized momentum $\mathbf{P}$ and the generalized energy $H$ of the particle in an electromagnetic field and the particle energy $E$ and momentum $\mathbf{p}$ can be represented in the form

$$P = \frac{\partial L}{\partial u} = \frac{mc_0 u}{\sqrt{1-u^2}} + \frac{e}{c_0}A = p + \frac{e}{c_0}A. \qquad H = P\cdot u - L = \frac{mc_0 c + e\varphi}{c_0} = \frac{E + e\varphi}{c_0}. \quad (5)$$

$$E = mc_0 c = mc_0^2\sqrt{1+\frac{v^2}{c_0^2}}, \qquad \mathbf{p}=m\mathbf{v}. \quad (6)$$

The particle momentum $\mathbf{p}$ and energy $E$ can be combined into the 4-momentum $p^\mu = mc_0 u^\mu = (mc_0 c/c_0, mcu^i) = (E/c_0, m\mathbf{v})$, $\mu = 0,1,2,3$, whose components are related by the equations $p_\mu p^\mu = E^2/c_0^2 - \mathbf{p}^2 = m^2 c_0^2$, $\mathbf{p} = (E/c_0 c)\mathbf{v}$, and $\mathbf{p} = (E/c_0 c)\mathbf{c}$. The last expression is valid at $m = 0$ and $\mathbf{v} = \mathbf{c}$. From here, it can be seen that the momentum of a particle having zero mass, $m = 0$, is independent of the absolute value of its velocity $v = c$ and is determined exclusively by its energy, $p = E/c_0$ and $\mathbf{p} = \mathbf{n}E/c_0$ with $\mathbf{n} = \mathbf{c}/c$, in just the same way as within special relativity [24]. For the purpose of illustration, the expressions for the momentum and energy, the relationship between them, and the equations of motion for a charged particle [21–23] are compiled in the table, where they are contrasted against the analogous relations within classical mechanics [1] and special relativity [24].

Comparison of space–time relations and basic properties of motion within classical mechanics, special relativity, and our study [1, 21–24]

|   | Classical mechanics | Special relativity | Our study |
|---|---|---|---|
| 1 | $ds^2 = dx^2 + dy^2 + dz^2$ | $ds^2 = c_0^2 dt^2 - dx^2 - dy^2 - dz^2$ | $ds^2 = (c_0^2 + v^2)dt^2 - dx^2 - dy^2 - dz^2$ |
| 2 | $x' = x - Ut$, $y' = y$, $z' = z$ <br> $t' = t$ <br> $c' = c[1-2(U/c)n_x+U^2/c^2]^{1/2}$ | $x' = (x - Wt)/(1 - ²)^{1/2}$, $y' = y$, <br> $z' = z$ <br> $t' = (t - xW/c_0^2)/(1 - W^2/c_0^2)^{1/2}$ <br> $c_0' = c_0 = 3 \times 10^{10}$ cm/s | $x' = (x - Vt)/(1 - ²)^{1/2}$, $y' = y$, <br> $z' = z$ <br> $t' = t$ <br> $c' = c(1 - V v_x/c^2)/(1 - V^2/c^2)^{1/2}$, |
| 3 | $u = U + u'$ | $W = (W + w')/(1 + Ww'/c_0^2)$ | $v' = (V + v'c/c')/(1 + Vv'/cc')$ |
| 4 | $\mathbf{p} = m\mathbf{u} = 2T\mathbf{u}/u^2$ | $\mathbf{p} = m\mathbf{w}/(1 - w^2/c_0^2)^{1/2} = E\mathbf{w}/c_0^2$ | $\mathbf{p} = m\mathbf{v} = E\mathbf{v}/c_0 c$ |
| 5 | $T = mu^2/2$ | $E = mc_0^2/(1 - w^2/c_0^2)^{1/2}$ | $E = mc_0^2(1 + v^2/c_0^2)^{1/2}$ |
| 6 | $T = p^2/2m$ | $E^2 - c_0^2 p^2 = m^2 c_0^4$ | $E^2 - c_0^2 p^2 = m^2 c_0^4$ |
| 7 | $m\dfrac{d\mathbf{u}}{dt} = e\mathbf{E} + \dfrac{e}{c_0}\mathbf{u} \times \mathbf{H}$ <br><br> $\dfrac{dT}{dt} = e\mathbf{u}\cdot\mathbf{E}$ | $\dfrac{d\mathbf{p}}{dt} = e\mathbf{E} + \dfrac{e}{c_0}\mathbf{w} \times \mathbf{H}$ <br><br> $\dfrac{dE}{dt} = e\mathbf{w}\cdot\mathbf{E}$ | $m\dfrac{d\mathbf{v}}{dt} = e\dfrac{c}{c_0}\mathbf{E} + \dfrac{e}{c_0}\mathbf{v} \times \mathbf{H}$ <br><br> $\dfrac{dE}{dt} = e\mathbf{v}\cdot\mathbf{E}$ |

Note: ($U$, $W$, $V$) velocity of the reference frame $K'$ with respect to $K$; (**u**, **w**, **v**) velocity of a particle with a mass $m$; (1) space–time metric, (2) space–time–velocity of light transformations, $n_x = \mathbf{U}\cdot\mathbf{c}/Uc$, = $W/c_0$ = $V/c$; (3) theorem of addition of velocities; (4) relation between the particle momentum **p** and the particle velocity (**u**, **w**, **v**) and energy $T$ and $E$; (5) relation between the energy and the particle velocity ($T$ is the kinetic energy); (6) relation between the energy and momentum; and (7) equations of motion for a charged particle (**E** and **H** are, respectively, the electric and magnetic fields). The Maxwell equations [21–23] have a standard form if the velocity of light $c_0$ in them is replaced by $c$ and if the electric-charge velocity **w** is replaced by **v**.

Let us now apply these relations to describing the kinematics of the Compton effect. Using the law of energy–momentum conservation, we write the set of equations that describe the scattering of a photon having an energy $E =$ on a free electron whose rest energy is $E_0 = mc_0^2$, where $m$ is the electron mass. We have [21–23]

$$\hbar\omega + E_0 = \hbar\omega' + \sqrt{1+v'^2/c_0^2}\,E_0, \quad \frac{\hbar\omega}{c_0} = \frac{\hbar\omega'}{c_0}\cos\theta + mv'\cos\alpha, \quad \frac{\hbar\omega'}{c_0}\sin\theta - mv'\sin\alpha = 0. \quad (7)$$

where $v'$ is the scattered-electron velocity and and are, respectively, the electron and photon scattering angles. For the velocity of a forward scattered electron ( = 0, = ), we obtain [21–23]

$$v'(\alpha=0, \theta=\pi) = c_0 \frac{\hbar\omega}{E_0} \frac{2(E_0 + \hbar\omega)}{E_0 + 2\hbar\omega}. \quad (8)$$

From (8), it follows that, at incident-photon energies in excess of $\omega > E_0/\sqrt{2} \sim 360$ keV, a faster than light motion of a Compton electron is possible, $v' > c_0$. This effect (if it exists) can be discovered by comparing, by means of the time-of-flight procedure, the velocity of an annihilation photon and the velocity of the Compton electron generated by it.

### 3. SCHEME OF THE PROPOSED EXPERIMENT

The layout of the facility for the proposed experiment is shown in the figure. The facility consists of

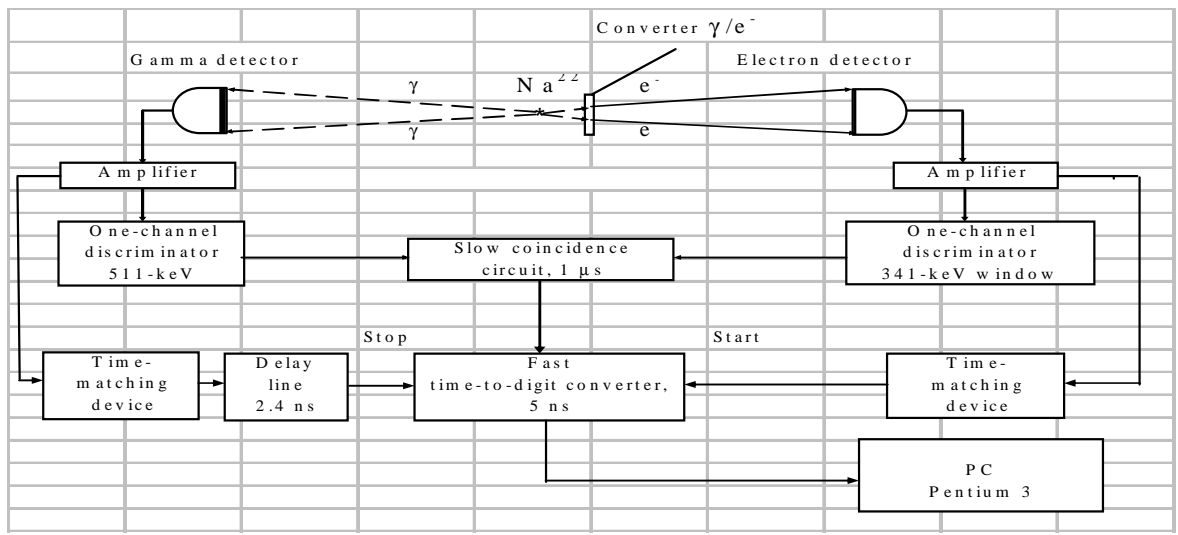

Layout of the facility for comparing the Compton electron and annihilation-photon velocities.

a $^{22}$Na($\beta^+$)$^{22}$Ne radioactive source; a vacuum chamber; a $\gamma/e^-$ converter; and two scintillator detectors, a $\gamma$ detector and an electron detector. The electronics of the facility relies on a fast–slow coincidence circuit including a slow channel for selecting photons and electrons in energy and a fast time channel for measuring time intervals between pulses from the two detectors. In the photon-detection channel, the differential-discriminator window is tuned to the annihilation-photon energy of 511 keV. In the electron channel, the window is tuned to the maximum kinetic energy of $T' = \{\gamma^2\beta^2(1 - \cos\theta)/[E_0 + \gamma(1 - \cos\theta)]\}_{\theta=\pi} = 2 \times 511/3 \approx 341$ keV, which is the energy of the total absorption of the Compton electron. At the time-channel width of 10 ps, a fast 5-ns time-to-digit converter corresponds to a 500-channel amplitude analyzer. The continuous spectrum of the distribution of photon times of flight with respect to the start pulse from the Compton electron is measured. If it turns out that the photon time of flight is shorter than the electron time of flight, then the $M^4_1$ is realized; in the opposite case, $M^4_2$ is realized.

### 4. ESTIMATING THE MAGNITUDE OF THE EFFECT

Suppose that the distance from the $^{22}$Na source to the detectors is 100 cm. The photon time of flight from the source to the detector is then about 3.3 ns. The calculated velocity of the Compton electron is about $0.8c_0$ in $M^4_1$ and $1.3c_0$ in $M^4_2$. The delay (outstripping) time between the electron and photon pulses is ±0.8 ns. We take the source activity to be 1 mCi = $3.7 \times 10^7$ decays per second and choose detectors based on plastic scintillators 120 mm in diameter. For a $\gamma/e^-$ converter, we imply a $(-CH_2-CH_2-)_n$ hydrocarbon film ($Z_{eff}/A_{eff} = 0.57$) of density $\rho = 0.7$ g/cm$^3$, $d \sim 0.01$ cm in thickness, and $S = 1$ cm$^2$ in area, the last value corresponding to the solid angle of the $\gamma$-detector acceptance. The efficiencies of the electron detector and the fast time-to-digit converter are taken to be 100%. The number of $\gamma-e^-$ coincidences is calculated by the formula $N_{\gamma-e} = \varepsilon_\gamma N_e$, where $\varepsilon_\gamma$ is the $\gamma$-detector efficiency and $N_e$ is the number of counts in the electron detector. According to [25], $\varepsilon_e = 18\%$ for an NE-104 organic scintillator 5 cm thick at the photon energy of 500 keV. The number of counts in the electron detector can be estimated by the formula $N_e = [A(\Omega_\gamma/4\pi)I_\gamma/S]\sigma_e n_e \varepsilon_e = 0.19$ pulse/s, where $A$ is the sodium-source activity, $\Omega_\gamma \approx \Omega_e = 1.1 \times 10^{-2}$ sr are the $\gamma$- and electron-detector solid angles, $I_\gamma = 0.9$ is the annihilation-line intensity (with allowance for absorption in the source) in the $\gamma$-detector direction, $\sigma_e = 23.6 \times 10^{-26}$ cm$^2/e$ sr is the cross section for Compton scattering of electrons at an angle of $\theta = 0$ into a unit solid angle, and $n_e = (Sd)(Z_{eff}/A_{eff})\rho N_0 = 2.4 \times 10^{21}$ is the number of scatterers (electrons) in the converter material ($N_0$ is Avogadro's number). The cross section for the Compton scattering of electrons into a solid angle $d\Omega$ is taken to be identical for the two Minkowski spaces on the basis of the formula for the differential scattering cross section [26]. This formula does not contain any identification features of the $M^4_1$ space, with the exception of the classical electron radius of $r_0 = e^2/E_0 = 2.82 \times 10^{-13}$ cm, which is identical for $M^4_1$ and $M^4_2$ since the electric charge $e$ of the electron and its rest energy $E_0$ are invariant. The expected number of coincidences is $N_{\gamma-e} \sim 0.18 \times 0.19 \sim 0.034$ pulse/s (about 120 pulse/h). As a measure of the noise, we take the number of random coincidences in the fast time-to-digit converter. If we increase the number of counts in the electron channel by a factor of ten because of the possible contribution of background electrons, then the number of random coincidences and the signal-to-noise ratio are $N_{ac} = 2N_\gamma N_e \tau \sim 1 \times 10^{-4}$ pulse/s and $\xi = N_{\gamma-e}/N_{ac}^{1/2} \sim 3.4$, respectively, where $N_\gamma = A(\Omega_\gamma/4\pi)I_\gamma\varepsilon_\gamma \sim 5.2 \times 10^3$ pulse/s in the 511-keV window and $\tau = 5$ ns is the converter time resolution. As a result, the useful signal exceeds the root-mean-square error of the noise by a factor of about 3.4, which is sufficient for detecting it against the background of random coincidences. The estimates obtained here are acceptable for performing the proposed experiment.

### 5. CONCLUSIONS

An experiment for testing special relativity on the basis of comparing the velocities of the Compton electron and the photon generating it has been proposed. The kinematics of the effect has been considered in the $M^4_2$ Minkowski space that belongs to the five-dimensional space of events also containing the $M^4_1$ Minkowski space (where special relativity is realized) and in which the invariance of the velocity of light is violated. In $M^4_2$, the velocity of light is isotropic by construction and the Lagrangian is invariant under a 12-dimensional group containing the Lorentz group as a subgroup. In the planned experiment, the annihilation-photon ($c_0 = 3 \times 10^{10}$ cm/s) time of flight is about 3.3 ns. If only the $M^4_1$ space is realized, then one peak must be observed experimentally in the time channel $n_1 = 3.3 + 2.4 - 3.3 - 0.8 = 1.6$ ns. If only the $M^4_2$ space is realized, one peak must be observed in the time channel $n_2 = (3.3 + 2.4 - 3.3 + 0.8) = 3.2$ ns. If the five-dimensional space containing the $M^4_1$ and $M^4_2$ subspaces is realized, the two peaks in question must be observed simultaneously. It should be noted that, in the history of testing special relativity, this is the first experiment where it is proposed to compare directly the velocity of a matter particle (Compton electron) and the velocity of a nonmatter particle (photon).

ACKNOWLEDGMENTS

This work was supported by the Russian Research Centre Kurchatov Institute.

*Translated by A. Isaakyan*